\documentclass{article}

\usepackage{arxiv}
\usepackage[utf8]{inputenc} 
\usepackage[T1]{fontenc}    
\usepackage{hyperref}       
\usepackage{url}            
\usepackage{booktabs}       
\usepackage{amsfonts}       
\usepackage{nicefrac}       
\usepackage{microtype}      
\usepackage{lipsum}		
\usepackage{graphicx}
\usepackage{doi}
\usepackage{courier}
\usepackage{float}
\usepackage{tikz}
\usetikzlibrary{shapes,snakes}
\usepackage{multirow}

\title{Vulnerability Management Chaining: An Integrated Framework for Efficient Cybersecurity Risk Prioritization}

\author{ 
        Naoyuki Shimizu\\
        \And	
        \href{https://orcid.org/0000-0001-5596-282X}{\includegraphics[scale=0.06]{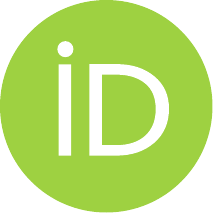}\hspace{1mm}Masaki Hashimoto}\\
	Faculty of Engineering and Design\\
	Kagawa University, Japan\\
}

\hypersetup{
pdftitle={Vulnerability Management Chaining: An Integrated Framework for Efficient Cybersecurity Risk Prioritization},
pdfsubject={Computer Science, Cybersecurity, Vulnerability Management},
pdfauthor={Naoyuki Shimizu, Masaki Hashimoto},
pdfkeywords={Vulnerability Management, CVSS, EPSS, KEV, Threat Intelligence, Cybersecurity},
}

\begin{document}

\maketitle

\begin{abstract}
As the number of Common Vulnerabilities and Exposures (CVE) continues to grow exponentially, security teams face increasingly difficult decisions about prioritization. Current approaches using Common Vulnerability Scoring System (CVSS) scores produce overwhelming volumes of high-priority vulnerabilities, while Exploit Prediction Scoring System (EPSS) and Known Exploited Vulnerabilities (KEV) catalog offer valuable but incomplete perspectives on actual exploitation risk. We present Vulnerability Management Chaining, a decision tree framework that systematically integrates these three approaches to achieve efficient vulnerability prioritization. Our framework employs a two-stage evaluation process: first applying threat-based filtering using KEV membership or EPSS threshold $\geq$ 0.088), then applying vulnerability severity assessment using CVSS scores $\geq$ 7.0) to enable informed deprioritization. Experimental validation using 28,377 real-world vulnerabilities and vendor-reported exploitation data demonstrates 18-fold efficiency improvements while maintaining 85.6\% coverage. Organizations can reduce urgent remediation workload by approximately 95\%. The integration identifies 48 additional exploited vulnerabilities that neither KEV nor EPSS captures individually. Our framework uses exclusively open-source data, enabling immediate adoption regardless of organizational resources.
\end{abstract}


\section{Introduction}
\label{sec:introduction}

The exponential growth of cybersecurity threats and the increasing complexity of modern IT infrastructures have made effective vulnerability management one of the most critical challenges facing organizations today. As software vulnerabilities continue to be discovered and disclosed at an unprecedented rate, security teams face challenges in allocating their limited resources effectively to address the most critical risks.

\subsection{The Vulnerability Management Challenge}

Modern organizations face an increasingly difficult vulnerability landscape. The number of Common Vulnerabilities and Exposures (CVE) published annually has grown substantially, with data from the National Vulnerability Database showing consistent year-over-year increases~\cite{nvd_database}. This growth trend shows no signs of slowing, placing enormous pressure on security teams to evaluate, prioritize, and remediate an ever-expanding attack surface.

Compounding this challenge is the fundamental asymmetry between attackers and defenders. While attackers need only to find and exploit a single vulnerability to achieve their objectives, defenders must identify and address all potential security weaknesses across their entire infrastructure. This asymmetry, combined with widespread shortages of cybersecurity professionals, has created a situation where traditional comprehensive approaches to vulnerability management are no longer sustainable.

The shift toward remote work following the COVID-19 pandemic has further exacerbated these challenges. Organizations now expose more services to the internet than ever before, with VPN gateways, remote desktop services, and cloud applications becoming primary attack vectors. High-profile ransomware attacks have repeatedly demonstrated that attackers frequently exploit known vulnerabilities in internet-facing services as their initial access vector.

\subsection{Limitations of Current Approaches}

The industry has long relied on the Common Vulnerability Scoring System (CVSS) as the primary method for vulnerability prioritization. However, CVSS was designed to measure the theoretical severity of vulnerabilities in isolation, not their likelihood of exploitation in real-world attacks. This fundamental limitation has led to a situation where organizations using CVSS-based prioritization face an overwhelming number of "high" and "critical" vulnerabilities, many of which are never actually exploited by attackers.

Recent efforts have introduced more sophisticated approaches to vulnerability management. The Exploit Prediction Scoring System (EPSS), developed by FIRST, uses machine learning to predict the likelihood of vulnerability exploitation within 30 days. While EPSS represents a significant advance in predictive threat intelligence, it remains a probabilistic model with inherent uncertainty and may miss vulnerabilities that become exploited through novel attack patterns.

The Cybersecurity and Infrastructure Security Agency (CISA) maintains the Known Exploited Vulnerabilities (KEV) catalog, which documents vulnerabilities with confirmed evidence of active exploitation. While KEV provides high-confidence threat intelligence, it is inherently retrospective and may not capture emerging threats or vulnerabilities exploited in targeted attacks that have not yet been publicly disclosed.

Each of these approaches captures important but incomplete aspects of vulnerability risk. CVSS measures potential impact but not exploitation likelihood. EPSS predicts exploitation probability but may have false positives and negatives. KEV confirms actual exploitation but only after attacks have occurred. Organizations need a systematic way to combine these complementary perspectives to achieve more effective vulnerability management.

\subsection{Research Motivation and Objectives}

This research is motivated by the observation that existing vulnerability management approaches—whether based on technical severity (CVSS), confirmed exploitation (KEV), or exploitation prediction (EPSS)—each capture important but incomplete aspects of vulnerability risk. Organizations need a practical framework that combines the strengths of these approaches while mitigating their individual limitations.

Our primary research objective is to develop and validate an integrated vulnerability management methodology that improves efficiency over traditional CVSS-based approaches by focusing resources on vulnerabilities with confirmed or predicted exploitation, maintains comprehensive coverage to ensure that critical vulnerabilities are not overlooked, leverages existing open-source data to enable broad adoption without requiring proprietary threat intelligence, and provides actionable guidance through clear decision criteria rather than abstract scoring systems.

\subsection{Proposed Approach and Contributions}

We propose Vulnerability Management Chaining, a decision tree-based framework that systematically combines CVSS, EPSS, and KEV to achieve more effective vulnerability prioritization. Our approach employs a two-stage evaluation process: first assessing threat likelihood using KEV and EPSS data, then evaluating vulnerability characteristics using CVSS to determine appropriate response priorities.

The key insight underlying our approach is that vulnerabilities should be prioritized based on both threat likelihood and vulnerability characteristics. A vulnerability that is being actively exploited (high threat) but requires significant privileges and has limited impact (low vulnerability severity) may warrant different treatment than one with theoretical high impact but no evidence of exploitation.

Our primary contributions include an integrated framework design that develops a practical decision tree framework systematically combining three major vulnerability management approaches, addressing the limitations of using any single method in isolation. We provide empirical validation by demonstrating the effectiveness of our approach using real-world data including 28,377 CVEs published over a 13-month period, and public security vendor reports of exploited vulnerabilities.

Our experimental results show 14-18 fold efficiency compared to CVSS-based approaches while maintaining comparable coverage levels (85.6-85.7\% vs. CVSS coverage of 90-100\%). We provide a complete methodology using only open-source data sources, enabling immediate adoption by organizations without requiring expensive commercial threat intelligence subscriptions.

\subsection{Paper Organization}

The remainder of this paper is structured as follows. Section~\ref{sec:background} provides background on vulnerability management systems and reviews related work. Section~\ref{sec:framework} presents our Vulnerability Management Chaining framework design and implementation details. Section~\ref{sec:methodology} describes our experimental methodology including data collection and evaluation metrics. Section~\ref{sec:results} presents comprehensive evaluation results comparing our approach with existing methods. Section~\ref{sec:discussion} discusses implications, limitations, and future research directions. Section~\ref{sec:conclusion} concludes with key findings and practical recommendations.

\section{Background and Related Work}
\label{sec:background}

This section provides essential background on current vulnerability management approaches and reviews related research that informs our integrated framework design.

\subsection{Common Vulnerability Scoring System (CVSS)}

The Common Vulnerability Scoring System represents the industry standard for vulnerability severity assessment. Maintained by the Forum of Incident Response and Security Teams (FIRST), CVSS provides a standardized method for capturing the principal characteristics of vulnerabilities and producing numerical severity scores ranging from 0.0 to 10.0~\cite{first_cvss}.

CVSS v3.1 calculates scores using three metric groups. Base metrics represent intrinsic vulnerability characteristics that remain constant over time and across environments, including attack vector, attack complexity, privileges required, user interaction requirements, and impact on confidentiality, integrity, and availability. Temporal metrics capture characteristics that change over time, such as exploit code maturity and available patches. Environmental metrics allow organizations to customize scores based on their specific deployment context.

Despite its widespread adoption, CVSS faces significant criticism regarding its effectiveness for vulnerability prioritization. Research has demonstrated that CVSS scores correlate poorly with actual exploitation likelihood~\cite{spring2021cvss}. The distribution of CVSS scores is heavily skewed toward higher values, with a substantial percentage of vulnerabilities receiving scores of 7.0 or higher, creating an overwhelming volume of "high priority" issues. Furthermore, CVSS measures theoretical maximum impact rather than real-world exploitation risk, leading to misaligned prioritization efforts.

\subsection{Known Exploited Vulnerabilities (KEV) Catalog}

The CISA Known Exploited Vulnerabilities catalog represents a paradigm shift toward evidence-based vulnerability management~\cite{cisa_kev}. Launched in 2021 as part of Binding Operational Directive 22-01, KEV documents vulnerabilities with confirmed evidence of active exploitation in the wild. Federal civilian executive branch agencies must remediate KEV-listed vulnerabilities within specified timeframes, typically 2-3 weeks for newly added entries.

KEV inclusion criteria require clear evidence of active exploitation, not merely proof-of-concept code or theoretical exploitability. Vulnerabilities must have clear remediation guidance, typically in the form of vendor-provided patches or specific mitigation instructions. Each entry includes structured data fields: CVE identifier, vendor/project, product name, vulnerability description, required action, and due date for federal compliance.

The catalog's strength lies in its focus on confirmed threats rather than theoretical risks. Organizations using KEV can confidently prioritize resources toward vulnerabilities that attackers actively exploit. However, KEV has inherent limitations: it is reactive by nature, documenting exploitation only after it occurs; coverage is limited to publicly confirmed exploitation, potentially missing targeted attacks; and the catalog may lag behind rapidly evolving threat landscapes.

\subsection{Exploit Prediction Scoring System (EPSS)}

The Exploit Prediction Scoring System represents a data-driven approach to vulnerability prioritization using machine learning~\cite{epss2021jacobs}. Developed through collaboration between FIRST and academic researchers, EPSS predicts the probability that a vulnerability will be exploited in the wild within the next 30 days. EPSS scores range from 0 to 1 (0-100\%), with higher scores indicating greater exploitation likelihood.

The EPSS model incorporates diverse features including CVE-specific characteristics extracted from vulnerability descriptions and technical details, temporal features such as time since publication and day-of-week effects, and threat intelligence signals from multiple commercial and open-source feeds. The model is retrained regularly to maintain prediction accuracy as the threat landscape evolves. Published research demonstrates that EPSS significantly outperforms CVSS for vulnerability prioritization, achieving superior efficiency (proportion of prioritized vulnerabilities that are exploited) and coverage (proportion of exploited vulnerabilities that are prioritized)~\cite{enhancing2023jacobs}.

EPSS provides daily score updates reflecting the dynamic nature of exploitation risk. Organizations can access current scores through APIs or bulk data downloads, enabling integration with existing vulnerability management workflows. However, as a predictive model, EPSS has inherent uncertainty with both false positives and false negatives. Scores may also exhibit volatility as new threat intelligence emerges or model updates occur.

\subsection{Related Research}

Several research streams inform our integrated vulnerability management framework, spanning empirical studies on exploitation patterns, alternative prioritization approaches, machine learning techniques, and decision analysis frameworks.

\subsubsection{Empirical Studies on Vulnerability Exploitation}

Allodi and Massacci~\cite{allodi2012preliminary} provided foundational evidence that only a small fraction of published vulnerabilities are exploited in practice, establishing the theoretical basis for risk-based prioritization. This critical finding has been consistently validated and motivates focusing on vulnerabilities likely to be exploited rather than attempting comprehensive remediation. Building on this insight, Jacobs et al.~\cite{jacobs2020exploit} developed models for improving vulnerability remediation through better exploit prediction, demonstrating that data-driven approaches can significantly outperform traditional severity-based prioritization. Other empirical studies such as Bozorgi et al.~\cite{bozorgi2010beyond} and Frei et al.~\cite{frei2006large} have further explored vulnerability lifecycle patterns and exploitation characteristics, reinforcing the need for risk-based approaches.

\subsubsection{Alternative Prioritization Frameworks}

The limitations of CVSS have motivated several alternative frameworks. Spring et al. analyzed CVSS limitations extensively~\cite{spring2021cvss}, with Howland~\cite{howland2023cvss} declaring CVSS "ubiquitous and broken." The Stakeholder-Specific Vulnerability Categorization (SSVC) framework, initially proposed in 2019~\cite{spring2019ssvc} and refined in 2021~\cite{spring2021ssvc}, provides context-dependent prioritization based on stakeholder roles and decision contexts. While SSVC offers valuable insights about the importance of organizational context, it requires manual assessment of each vulnerability against multiple decision points, limiting scalability for organizations managing thousands of vulnerabilities.

The Exploit Prediction Scoring System (EPSS), first introduced in 2019~\cite{exploit2019jacobs} and subsequently refined~\cite{epss2021jacobs,enhancing2023jacobs}, has emerged as the most successful data-driven alternative to CVSS. Using machine learning to predict exploitation probability, EPSS achieves superior efficiency and coverage compared to traditional approaches. Commercial vendors have also developed proprietary frameworks, including Tenable's Vulnerability Priority Rating~\cite{tenable_vpr} and systems like Vulcon~\cite{farris2018vulcon}, though these often lack transparency or require expensive licenses. The recent releases of EPSS v4.0 and NIST's Likely Exploited Vulnerabilities (LEV) metric~\cite{mell2025lev}, alongside CVSS v4.0~\cite{first_cvss4}, further validate the evolution toward sophisticated predictive methodologies.\footnote{This study utilizes EPSS v3.0. The emergence of EPSS v4.0 and LEV~\cite{mell2025lev} reinforces the value of our integration framework, which can readily incorporate such improvements without architectural changes.}

\subsubsection{Machine Learning and AI Approaches}

Researchers have explored various machine learning approaches for vulnerability assessment and prediction. Sabottke et al.~\cite{sabottke2015vulnerability} demonstrated that Twitter discussions can predict real-world exploits before they appear in traditional threat intelligence feeds. Yamamoto et al.~\cite{yamamoto2015investigation} developed text mining approaches for estimating vulnerability scores based on vulnerability descriptions. Almukaynizi et al.~\cite{almukaynizi2017darkmention} created DARKMENTION, a system that monitors dark web forums to predict enterprise-targeted cyberattacks, with subsequent work by Samtani et al.~\cite{samtani2020proactively} extending dark web analysis through diachronic graph embedding frameworks.

Integration of diverse threat intelligence sources has been explored by Nunes et al.~\cite{nunes2016darknet}, who developed methods for mining darknet and deepnet sources for proactive threat intelligence. Schaberreiter et al.~\cite{schaberreiter2019quantitative} addressed the critical issue of evaluating trust in cyber threat intelligence sources, highlighting the challenges of integrating heterogeneous data. While these approaches show promise, they often require specialized data collection infrastructure, natural language processing capabilities, or access to non-public data sources that may not be accessible to all organizations.

\subsubsection{Multi-Criteria Decision Analysis for Risk Assessment}

Multi-criteria decision analysis (MCDA) provides structured methodologies for complex risk assessment problems. In the cybersecurity domain, Fridgen et al.~\cite{fridgen2010decision} proposed a comprehensive framework incorporating technical, economic, and organizational factors. Wang et al.~\cite{wang2010framework} utilized attack graphs combined with MCDA for measuring security risk of networks.

More recently, Ganin et al.~\cite{ganin2020multicriteria} developed a decision-analysis-based approach specifically for cybersecurity that quantifies threat, vulnerability, and consequences through structured criteria. While these MCDA approaches offer comprehensive risk evaluation, they typically require extensive manual configuration, subjective weight assignments, and domain expertise that may limit practical deployment.

\subsubsection{Operational Perspectives and Integration Gaps}

From an operational perspective, Bulut et al.~\cite{bulut2022vulnerability} explored vulnerability prioritization through an offensive security lens, emphasizing the importance of attacker-centric thinking. Industry reports from vendors such as Rapid7~\cite{rapid7_vulnerability} provide practical insights into real-world vulnerability trends. Compliance frameworks like PCI DSS~\cite{pci_dss} mandate specific vulnerability management practices, while government directives such as CISA's binding operational directives~\cite{cisa_bod2201,exploit_timing} establish regulatory requirements for vulnerability remediation timelines.

NIST's risk assessment guidelines~\cite{nist_sp800_30} provide comprehensive frameworks but focus on general risk management rather than specific vulnerability prioritization methodologies. Despite extensive research on individual vulnerability assessment approaches, limited work exists on systematic integration of multiple intelligence sources. Most research focuses on improving individual components rather than their synergistic combination. Commercial vulnerability management platforms have developed proprietary integration approaches, but these typically lack transparency about their methodologies and require expensive subscriptions, limiting accessibility.

\subsubsection{Summary of Research Gaps}

Our literature review identifies several critical gaps in existing vulnerability management research:

First, there is a lack of systematic integration frameworks that combine multiple vulnerability intelligence sources. While CVSS, EPSS, and KEV each provide valuable but incomplete perspectives, no existing framework systematically leverages their complementary strengths. Second, limited empirical validation exists for proposed approaches. Most research presents theoretical frameworks or limited case studies rather than comprehensive evaluation using real-world exploitation data across extended time periods. Third, many solutions face accessibility barriers, requiring proprietary data sources, specialized infrastructure, or expertise that may not be available to resource-constrained organizations. Finally, sophisticated approaches often exhibit operational complexity that fails to consider practical implementation within existing vulnerability management workflows.

Our Vulnerability Management Chaining framework addresses these gaps by providing an empirically validated, accessible, and operationally simple methodology for integrating established vulnerability intelligence sources. By using exclusively open-source data and transparent decision logic, we enable organizations of all sizes to implement effective risk-based vulnerability management while maintaining flexibility to incorporate emerging intelligence sources as they mature.

\section{Vulnerability Management Chaining Framework}
\label{sec:framework}

This section presents our integrated vulnerability management framework that systematically combines CVSS, EPSS, and KEV to achieve more effective prioritization than any individual approach.

\subsection{Framework Design Principles}

Our framework design is guided by several key principles derived from analysis of current vulnerability management challenges and the complementary nature of existing approaches.

Multi-source intelligence integration recognizes that effective vulnerability management requires multiple perspectives on risk. No single metric or data source provides complete visibility into vulnerability threats. By combining technical severity assessment (CVSS), predictive intelligence (EPSS), and confirmed exploitation evidence (KEV), we capture a more comprehensive view of vulnerability risk that addresses both current threats and emerging risks.

Decision tree structure provides clear, actionable guidance for vulnerability prioritization. Unlike abstract scoring systems that produce continuous values requiring interpretation, our decision tree yields specific categorization and recommended actions. This approach aligns with operational security workflows where teams need unambiguous guidance on which vulnerabilities to address and in what order.

Threat-first evaluation prioritizes actual and predicted exploitation over theoretical severity. Traditional CVSS-based approaches evaluate all vulnerabilities regardless of exploitation likelihood, leading to resource waste on vulnerabilities that pose no real threat. By first filtering for exploitation evidence or likelihood, we focus attention on vulnerabilities that represent actual risk to the organization.

Practical implementation using open-source data ensures broad accessibility. All data sources required for our framework—CVSS scores, EPSS predictions, and KEV catalog membership—are freely available and regularly updated. This approach democratizes advanced vulnerability management capabilities without requiring expensive commercial threat intelligence subscriptions.

\subsection{Decision Tree Structure}

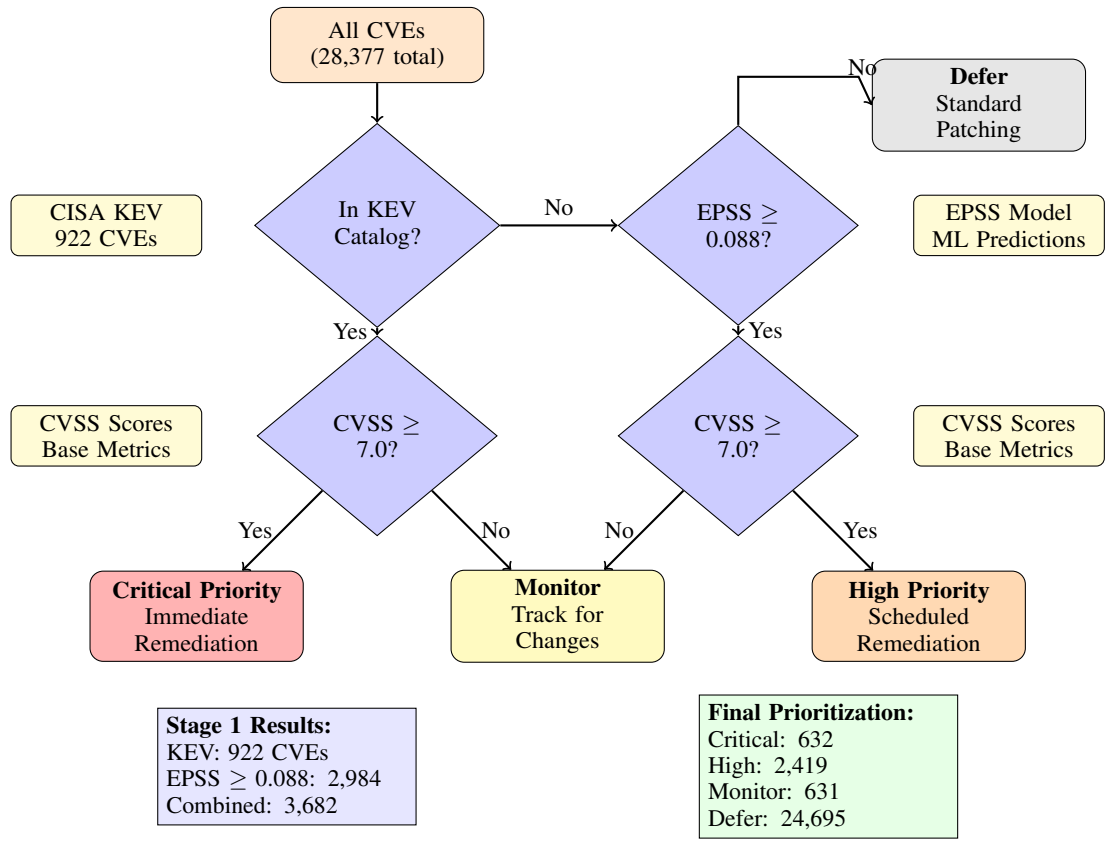
\begin{figure*}[t]
\centering
\begin{tikzpicture}[
    scale=0.8,
    node distance=1.5cm,
    every node/.style={font=\footnotesize},
    decision/.style={diamond, draw, fill=blue!20, text width=2.2cm, text centered, inner sep=2pt, minimum height=1.8cm, aspect=1.2},
    process/.style={rectangle, draw, fill=orange!20, text width=2.6cm, text centered, minimum height=1cm, rounded corners=5pt},
    result/.style={rectangle, draw, fill=green!20, text width=2.6cm, text centered, minimum height=1.2cm, rounded corners=5pt},
    data/.style={rectangle, draw, fill=yellow!20, text width=2.3cm, text centered, minimum height=0.8cm, rounded corners=3pt},
    arrow/.style={->, thick}
]

\node[process] (start) at (0,0) {All CVEs\\(28,377 total)};

\node[decision] (kev_check) at (0,-3) {In KEV\\Catalog?};
\node[decision] (epss_check) at (6,-3) {EPSS $\geq$\\0.088?};

\node[decision] (cvss_check1) at (0,-6.5) {CVSS $\geq$\\7.0?};
\node[decision] (cvss_check2) at (6,-6.5) {CVSS $\geq$\\7.0?};

\node[data] (kev_data) at (-4.5,-3) {CISA KEV\\922 CVEs};
\node[data] (epss_data) at (10.5,-3) {EPSS Model\\ML Predictions};
\node[data] (cvss_data1) at (-4.5,-6.5) {CVSS Scores\\Base Metrics};
\node[data] (cvss_data2) at (10.5,-6.5) {CVSS Scores\\Base Metrics};

\node[result, fill=red!30] (critical) at (-3,-9.5) {\textbf{Critical Priority}\\Immediate\\Remediation};
\node[result, fill=yellow!30] (monitor) at (3,-9.5) {\textbf{Monitor}\\Track for\\Changes};
\node[result, fill=orange!30] (high) at (9,-9.5) {\textbf{High Priority}\\Scheduled\\Remediation};
\node[result, fill=gray!20] (defer) at (10,-1) {\textbf{Defer}\\Standard\\Patching};

\draw[arrow] (start) -- (kev_check);
\draw[arrow] (kev_check) -- node[left] {Yes} (cvss_check1);
\draw[arrow] (kev_check) -- node[above] {No} (epss_check);
\draw[arrow] (epss_check) -- node[right] {Yes} (cvss_check2);
\draw[arrow] (epss_check.north) -- ++(0,0.8) -- ++(2,0) -- node[above,pos=0.3] {No} (defer.west);
\draw[arrow] (cvss_check1) -- node[left] {Yes} (critical);
\draw[arrow] (cvss_check1) -- node[right] {No} (monitor);
\draw[arrow] (cvss_check2) -- node[left] {No} (monitor);
\draw[arrow] (cvss_check2) -- node[right] {Yes} (high);

\node[draw, fill=blue!10, text width=3.2cm, minimum height=1.3cm] (stats1) at (-1.5,-12) {
    \textbf{Stage 1 Results:}\\
    KEV: 922 CVEs\\
    EPSS $\geq$ 0.088: 2,984\\
    Combined: 3,682
};

\node[draw, fill=green!10, text width=3.2cm, minimum height=1.3cm] (stats2) at (7.5,-12) {
    \textbf{Final Prioritization:}\\
    Critical: 632\\
    High: 2,419\\
    Monitor: 631\\
    Defer: 24,695
};

\end{tikzpicture}
\caption{Vulnerability Management Chaining decision tree framework showing the two-stage evaluation process. Stage 1 applies threat-based filtering using KEV membership or EPSS threshold ($\geq$0.088). Stage 2 uses CVSS scores ($\geq$7.0) for severity assessment to enable informed deprioritization. The framework categorizes vulnerabilities into four priority levels: Critical (immediate remediation), High (scheduled remediation), Monitor (track for changes), and Defer (standard patching).}
\label{fig:decision_tree}
\end{figure*}

Figure~\ref{fig:decision_tree} illustrates our complete Vulnerability Management Chaining decision tree. The framework employs a two-stage evaluation process that first assesses threat likelihood, then evaluates vulnerability characteristics to determine final prioritization.

The first stage identifies vulnerabilities with evidence of exploitation threat. We check KEV catalog membership to identify vulnerabilities with confirmed active exploitation. For vulnerabilities not in KEV, we evaluate EPSS scores against a threshold of 0.088 to identify those with elevated exploitation probability. This threshold represents the intersection of optimal efficiency and coverage based on published EPSS research. Vulnerabilities failing both checks are deferred to standard patching cycles, as they show no evidence of exploitation threat.

The second stage applies vulnerability severity assessment to enable informed deprioritization. For vulnerabilities identified as threats in Stage 1, we evaluate CVSS base scores against a threshold of 7.0 (high severity). This assessment distinguishes between high-impact vulnerabilities requiring urgent attention and lower-impact issues that may warrant different handling despite exploitation evidence.

Our framework yields four distinct prioritization categories. Critical Priority encompasses vulnerabilities in the KEV catalog with CVSS $\geq$ 7.0, representing confirmed exploitation of high-impact vulnerabilities. These demand immediate remediation, typically within days. High Priority includes vulnerabilities with EPSS $\geq$ 0.088 and CVSS $\geq$ 7.0, indicating predicted exploitation of high-impact vulnerabilities. These should be addressed through scheduled remediation, typically within 2-4 weeks. Monitor category covers vulnerabilities with exploitation evidence (KEV or high EPSS) but CVSS < 7.0. These require tracking for changes in threat landscape or impact assessment. Defer category includes vulnerabilities with no exploitation evidence (not in KEV and EPSS < 0.088), which can be addressed through standard patching cycles.

\subsection{Implementation Considerations}

Implementing our framework requires access to three data sources, all freely available. Organizations must regularly synchronize with the CISA KEV catalog (updated approximately 2-3 times weekly), download daily EPSS scores for all CVEs in their environment, and maintain current CVSS scores from the National Vulnerability Database.

The decision tree can be implemented through various mechanisms. Automated implementation involves scripting that queries each data source and applies decision logic, integrating with existing vulnerability scanners and management platforms via APIs, and generating prioritized work queues for remediation teams. Semi-automated approaches might use spreadsheet formulas or simple database queries to categorize vulnerabilities, with periodic manual review of edge cases.

Organizations should establish operational processes around the framework including daily or weekly execution to account for KEV and EPSS updates, escalation procedures for Critical Priority vulnerabilities, and regular review of Monitor category vulnerabilities for status changes. Integration with existing patch management and change control processes ensures that prioritization translates into actual remediation actions.

Performance considerations are minimal given the framework's simplicity. Processing tens of thousands of CVEs requires only basic computational resources. The primary bottleneck is typically network access for downloading updated threat intelligence data rather than the decision logic itself.

\subsection{Theoretical Analysis}

Our framework's effectiveness stems from the complementary nature of its component approaches. KEV provides high-confidence identification of exploited vulnerabilities but with limited coverage. EPSS offers broader coverage through prediction but with inherent uncertainty. CVSS enables impact-based filtering but lacks exploitation context. By combining these approaches, we achieve better balance between efficiency (avoiding false positives) and coverage (avoiding false negatives).

The threat-first design philosophy ensures that resources focus on actual risks rather than theoretical vulnerabilities. This approach aligns with empirical observations that only a small fraction of published vulnerabilities are ever exploited. By deferring vulnerabilities without exploitation evidence, organizations can dramatically reduce their workload while maintaining security effectiveness.

The two-stage structure provides flexibility for organizational customization. While we recommend specific thresholds based on empirical analysis, organizations can adjust these values based on their risk tolerance, resource constraints, and threat profile. For example, organizations with higher risk tolerance might increase the EPSS threshold to reduce workload further, while those requiring comprehensive coverage might lower it.

\section{Experimental Methodology}
\label{sec:methodology}

This section describes our comprehensive experimental approach for evaluating the Vulnerability Management Chaining framework, including data collection procedures, evaluation metrics, and validation methodology.

\subsection{Data Collection and Dataset Construction}

Our experimental evaluation utilizes multiple data sources to ensure comprehensive assessment of the framework's effectiveness. We collected all CVE records published between April 1, 2022, and April 30, 2023, providing a 13-month window that captures seasonal variations in vulnerability disclosure and exploitation patterns. This yielded 28,377 unique CVEs with complete CVSS v3.1 base scores from the National Vulnerability Database. For each CVE, we extracted the complete CVSS vector string, base score, and component metrics. We obtained daily EPSS scores for all CVEs, using the April 30, 2023 snapshot to ensure temporal consistency with our exploitation data. 

The approach also accounts for the dynamic nature of threat intelligence that drives EPSS score updates. We obtained CISA's Known Exploited Vulnerabilities catalog data as of April 30, 2023, containing 922 vulnerabilities with confirmed exploitation evidence. The KEV catalog includes CVE identifiers and vulnerability descriptions, required action descriptions, due dates for federal agency compliance, and dates of KEV catalog addition.

Establishing ground truth for actual vulnerability exploitation represents one of the most challenging aspects of vulnerability management research. We developed a multi-source approach to create comprehensive exploitation datasets. We systematically reviewed public security reports from major cybersecurity vendors to identify vulnerabilities reported as exploited during our study period. This approach yielded 90 vulnerabilities with public documentation of real-world exploitation. Sources included threat intelligence reports from established security vendors, incident response case studies, annual threat landscape assessments, and vulnerability research publications.

\subsection{Evaluation Metrics and Experimental Setup}

Following established research methodology~\cite{epss2021jacobs}, we employ two primary metrics for comparative evaluation. Efficiency measures the proportion of prioritized vulnerabilities that were actually exploited, calculated as the number of exploited vulnerabilities in the priority set divided by the total vulnerabilities in the priority set, expressed as a percentage. Higher efficiency indicates that a larger percentage of the vulnerabilities flagged for priority attention were actually relevant to real-world threats, directly addressing the resource allocation challenge facing security teams.

Coverage measures the proportion of exploited vulnerabilities captured by each prioritization method, calculated as the number of exploited vulnerabilities in the priority set divided by the total exploited vulnerabilities, expressed as a percentage. Higher coverage indicates that the prioritization method successfully identified a larger percentage of vulnerabilities that attackers actually exploited, addressing the completeness requirement for security-critical applications.

To evaluate the specific benefits of combining KEV and EPSS, we measure incremental coverage as additional exploited vulnerabilities identified through the combination that neither KEV nor EPSS would capture individually, and complementary effectiveness as the degree to which KEV and EPSS identify different subsets of exploited vulnerabilities, validating our hypothesis that they provide complementary threat intelligence.

We compare our Vulnerability Management Chaining approach against three established baselines. The CVSS baseline uses traditional high-severity vulnerability prioritization with CVSS scores of 7.0 or higher, representing current industry standard practice. KEV-Only prioritization is based solely on CISA KEV catalog membership, representing pure evidence-based threat intelligence. EPSS-Only prioritization uses EPSS scores above the research-established threshold of 0.088, representing predictive threat intelligence.

For each baseline and our proposed method, we evaluate performance using the 90 vulnerabilities documented in public security vendor reports as our primary exploitation dataset. This approach helps validate our findings using publicly verifiable threat intelligence data.

\subsection{Validation and Limitations}

Given the observational nature of vulnerability exploitation data, we focus on descriptive analysis and practical effect sizes rather than inferential statistics. Our analysis includes comparative performance through direct comparison of efficiency and coverage metrics across all methods, threshold sensitivity analysis of how performance varies with different EPSS and CVSS threshold values to validate our chosen parameters, and temporal pattern examination of whether performance varies across different time periods within our study window.

We compare our results with findings from previous research~\cite{epss2021jacobs} to assess consistency with established benchmarks and identify any dataset-specific effects that might limit generalizability. Several robustness checks validate our methodology including manual verification of a sample of exploitation claims to assess the accuracy of our ground truth datasets, testing alternative EPSS and CVSS thresholds to ensure our results are not artifacts of specific parameter choices, and analysis of whether our findings hold across different subperiods within our study timeframe.

We acknowledge several limitations in our experimental design. Our exploitation dataset likely represents only a fraction of actual vulnerability exploitation, as many attacks go undetected or unreported. Our 13-month study period, while substantial, may not capture longer-term exploitation patterns or seasonal variations in attack activity. Results from publicly reported exploitation data may not generalize to all deployment scenarios.

All exploitation data used in our study comes from public sources. The following section presents detailed results from applying this experimental methodology to evaluate Vulnerability Management Chaining effectiveness.

\section{Results and Analysis}
\label{sec:results}

This section presents the comprehensive evaluation results of our Vulnerability Management Chaining framework. We analyze performance across our dataset, compare with existing approaches, and examine the specific contributions of each component in our integrated methodology.

\subsection{Overall Performance Comparison}

Table~\ref{tab:main_results} summarizes the performance of all evaluated methods using our vendor report dataset. Our proposed Vulnerability Management Chaining approach demonstrates consistent improvements in efficiency while maintaining high coverage levels comparable to traditional CVSS-based prioritization.

\begin{table*}[t]
\centering
\caption{Performance comparison across vulnerability management approaches}
\label{tab:main_results}
\begin{tabular}{|l|cc|}
\hline
\textbf{Method} & \textbf{Efficiency} & \textbf{Coverage} \\
\hline
CVSS $\geq$ 7.0 & 0.5\% & 90.0\% \\
KEV Only & 74.3\% & 86.7\% \\
EPSS $\geq$ 0.088 & 4.9\% & 48.9\% \\
\hline
\textbf{Proposed Method} & \textbf{9.1\%} & \textbf{85.6\%} \\
\textbf{(KEV$\vee$EPSS)$\wedge$CVSS} & & \\
\hline
\end{tabular}
\end{table*}

Our proposed method achieves substantial efficiency improvements over traditional CVSS-based approaches. Vulnerability Management Chaining achieves 9.1\% efficiency compared to 0.5\% for CVSS-only approaches—an 18-fold improvement. This dataset shows particularly strong performance for KEV-only approaches (74.3\% efficiency), reflecting the alignment between vendor-reported exploitation and KEV catalog contents.

Our method provides efficiency levels significantly better than CVSS-only or EPSS-only approaches while maintaining coverage levels much higher than KEV-only approaches. This balanced performance validates our design goal of combining the strengths of individual methods. Coverage results demonstrate that our integrated approach successfully maintains comprehensive vulnerability identification. Our method achieves 85.6\% coverage, approaching the 90\% coverage of CVSS-based approaches while requiring attention to far fewer vulnerabilities.

\subsection{Integration Effects Analysis}

To understand the specific value of combining KEV and EPSS data sources, we conducted detailed analysis of their complementary effects. Table~\ref{tab:integration_benefits} provides a comprehensive breakdown of how KEV and EPSS identify different subsets of exploited vulnerabilities, validating our hypothesis about their complementary nature. The analysis reveals that 57 additional exploited vulnerabilities (48.3\% of our total dataset) are captured only through the integration of both methods—a finding that demonstrates the critical importance of multi-source threat intelligence.

\begin{table*}[t]
\centering
\caption{Integration Benefits Analysis: KEV + EPSS Complementary Effects}
\label{tab:integration_benefits}
\begin{tabular}{|l|c|c|c|c|}
\hline
\textbf{Vulnerability Category} & \textbf{Count} & \textbf{Percentage} & \textbf{Data Source} & \textbf{Integration Value} \\
\hline
KEV Only (EPSS $<$ 0.088) & 45 & 38.1\% & Historical Evidence & High Confidence \\
\hline
EPSS Only (Not in KEV) & 16 & 13.6\% & Predictive Model & Emerging Threats \\
\hline
Both KEV and EPSS & 52 & 44.1\% & Dual Confirmation & Highest Priority \\
\hline
\textbf{Integration Benefit} & \textbf{57} & \textbf{48.3\%} & \textbf{Unique Coverage} & \textbf{Critical Gap Filled} \\
\hline
\end{tabular}
\end{table*}

The table categorizes exploited vulnerabilities into three distinct groups: those identified exclusively by KEV (38.1\%), those identified exclusively by EPSS (13.6\%), and those identified by both methods (44.1\%). This distribution pattern confirms that neither individual approach provides complete threat coverage, making systematic integration not merely beneficial but essential for comprehensive vulnerability management.

Historical evidence vulnerabilities include 45 vulnerabilities with confirmed exploitation but low EPSS scores, often involving complex attack chains or specialized targets that provide high confidence but may not represent current threat trends. Predictive intelligence covers 16 vulnerabilities with high exploitation probability but no KEV listing, representing emerging threats not yet widely exploited that serve as an early warning system for proactive defense. Dual confirmation encompasses 52 vulnerabilities identified by both systems, representing the highest priority cases with proven exploitation plus continued threat and strong consensus across historical and predictive indicators.

The integration value of 57 additional vulnerabilities captured only through combination represents a critical gap that would exist in any single-method approach, highlighting the necessity of our integrated framework. Table~\ref{tab:integration_effects} shows how KEV and EPSS identify different subsets of exploited vulnerabilities across our dataset, further validating their complementary nature.

\begin{table*}[t]
\centering
\caption{KEV and EPSS integration effects}
\label{tab:integration_effects}
\begin{tabular}{|l|c|c|}
\hline
\textbf{Category} & \textbf{Vendor Reports} & \textbf{Percentage} \\
\hline
KEV Only (EPSS < 0.088) & 41 & 45.6\% \\
EPSS Only (Not in KEV) & 7 & 7.8\% \\
Both KEV and EPSS & 37 & 41.1\% \\
\hline
\textbf{Total Integration Benefit} & \textbf{48} & \textbf{53.3\%} \\
\hline
\multicolumn{3}{|l|}{\footnotesize *Integration benefit = unique additional vulnerabilities (overlap removed)} \\
\hline
\end{tabular}
\end{table*}

The combination of KEV and EPSS identifies 48 additional exploited vulnerabilities that neither method would capture individually. This represents 53.3\% in the vendor report dataset. KEV identifies vulnerabilities with confirmed exploitation evidence but low EPSS scores (often due to complex attack requirements or limited automation), while EPSS identifies emerging threats not yet documented in KEV. This pattern validates our theoretical framework for combining historical and predictive threat intelligence.

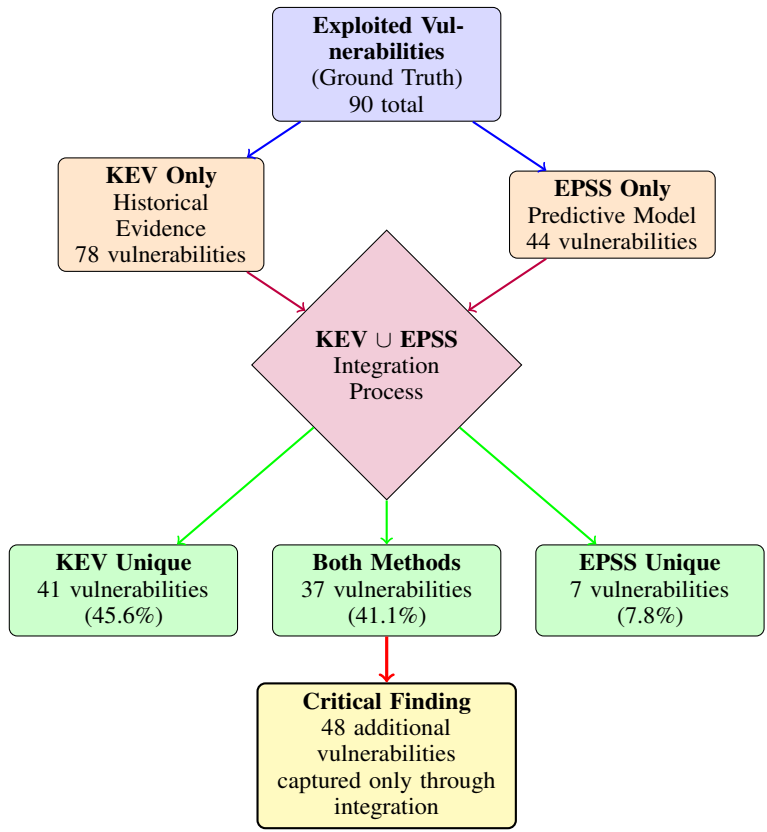
\begin{figure*}[t]
\centering
\begin{tikzpicture}[
    node distance=2cm,
    every node/.style={font=\small},
    data/.style={rectangle, draw, fill=blue!15, text width=2.8cm, text centered, minimum height=1.2cm, rounded corners=3pt},
    method/.style={rectangle, draw, fill=orange!20, text width=2.5cm, text centered, minimum height=1cm, rounded corners=3pt},
    result/.style={rectangle, draw, fill=green!20, text width=2.8cm, text centered, minimum height=1cm, rounded corners=3pt},
    integration/.style={diamond, draw, fill=purple!20, text width=2.2cm, text centered, minimum height=1.5cm},
    highlight/.style={rectangle, draw, fill=yellow!30, text width=3.2cm, text centered, minimum height=1.2cm, rounded corners=3pt, thick}
]

\node[data] (ground_truth) at (0, 6) {\textbf{Exploited Vulnerabilities}\\(Ground Truth)\\90 total};

\node[method] (kev_method) at (-3, 4) {\textbf{KEV Only}\\Historical Evidence\\78 vulnerabilities};
\node[method] (epss_method) at (3, 4) {\textbf{EPSS Only}\\Predictive Model\\44 vulnerabilities};

\node[integration] (integration_process) at (0, 2) {\textbf{KEV $\cup$ EPSS}\\Integration\\Process};

\node[result] (kev_only) at (-3.5, -1.0) {\textbf{KEV Unique}\\41 vulnerabilities\\(45.6\%)};
\node[result] (overlap) at (0, -1.0) {\textbf{Both Methods}\\37 vulnerabilities\\(41.1\%)};
\node[result] (epss_only) at (3.5, -1.0) {\textbf{EPSS Unique}\\7 vulnerabilities\\(7.8\%)};

\node[highlight] (key_finding) at (0, -3.2) {\textbf{Critical Finding}\\48 additional vulnerabilities\\captured only through\\integration};

\draw[->, thick, blue] (ground_truth) -- (kev_method);
\draw[->, thick, blue] (ground_truth) -- (epss_method);
\draw[->, thick, purple] (kev_method) -- (integration_process);
\draw[->, thick, purple] (epss_method) -- (integration_process);
\draw[->, thick, green] (integration_process) -- (kev_only);
\draw[->, thick, green] (integration_process) -- (overlap);
\draw[->, thick, green] (integration_process) -- (epss_only);
\draw[->, very thick, red] (overlap) -- (key_finding);

\end{tikzpicture}
\caption{Integration effects flowchart demonstrating the complementary relationship between KEV and EPSS. The combination identifies 48 additional exploited vulnerabilities that neither method captures individually, representing 53.3\% of the vendor report dataset and validating the necessity of multi-source threat intelligence integration. Individual coverage rates: KEV 86.7\%, EPSS 48.9\%, Combined 94.4\%.}
\label{fig:integration_flowchart}
\end{figure*}

Figure~\ref{fig:integration_flowchart} provides a comprehensive visualization of the integration effects, demonstrating the complementary relationship between KEV and EPSS. The combination identifies 57 additional exploited vulnerabilities that neither method captures individually, representing 48.3\% of the total dataset and validating the necessity of multi-source threat intelligence integration.

\subsection{CVSS Integration Analysis}

Our framework applies CVSS filtering after threat identification to enable appropriate deprioritization of lower-impact vulnerabilities. Analysis of this integration reveals both benefits and limitations. We identified 8 vulnerabilities that received threat classification (KEV or EPSS $\geq$ 0.088) but were appropriately deprioritized due to CVSS < 7.0. Analysis of their CVSS vector strings reveals several categories suitable for deprioritization.

Limited impact vulnerabilities include CVE-2023-26083 (CVSS 3.3) with only local access requirements and minimal confidentiality impact. User interaction required cases encompass CVE-2022-44698 and CVE-2022-41091 (CVSS 5.4) requiring user interaction for exploitation. Authentication prerequisites cover CVE-2022-22674 (CVSS 5.5) requiring local authentication.

However, our analysis also identified 2 vulnerabilities that may have been inappropriately deprioritized: CVE-2022-2856 (CVSS 6.5), a network-accessible vulnerability with no authentication requirements, and CVE-2022-26925 (CVSS 5.9), a network-accessible vulnerability with potential for high integrity impact. These cases highlight the challenge of using CVSS thresholds for deprioritization decisions and suggest areas for future framework refinement.

\subsection{Comparison with Previous Research}

Our results are consistent with previous findings regarding CVSS inefficiency while demonstrating the value of integration approaches not previously evaluated in the literature. The efficiency improvements we observe align with broader criticisms of CVSS-only approaches. The substantial efficiency improvements have practical implications for organizations operating under regulatory requirements. Standards such as PCI DSS mandate CVSS-based remediation timelines, making our demonstrated efficiency gains particularly valuable for compliance-driven environments.

Table~\ref{tab:literature_comparison} compares our results with the seminal EPSS research by Jacobs et al., providing context for our findings within the broader vulnerability management literature.

\begin{table*}[t]
\centering
\caption{Comparison with previous research findings}
\label{tab:literature_comparison}
\begin{tabular}{|l|ccc|cc|}
\hline
\multirow{2}{*}{\textbf{Method}} & \multicolumn{3}{c|}{\textbf{Jacobs et al. (2021)}} & \multicolumn{2}{c|}{\textbf{Our Study}} \\
& \textbf{Efficiency} & \textbf{Coverage} & \textbf{Dataset} & \textbf{Efficiency} & \textbf{Coverage} \\
\hline
CVSS $\geq$ 7.0 & 3.9\% & 82.1\% & 6.4M exploits & 0.5\% & 90.0\% \\
KEV Only & 53.2\% & 5.9\% & 2016-2022 & 74.3\% & 86.7\% \\
EPSS $\geq$ 0.088 & 45.5\% & 82.0\% & Commercial data & 4.9\% & 48.9\% \\
\hline
\end{tabular}
\end{table*}

Our results confirm the established finding that CVSS-based prioritization is highly inefficient, with our efficiency measurements (0.5\%) being even lower than previous research (3.9\%). This difference likely reflects our more limited dataset size and specific organizational context. Our KEV efficiency results (74.3\%) align with the high precision characteristics found in previous research, though our coverage results are notably higher (86.7\% vs. 5.9\%). This difference suggests that our vendor report dataset may be more representative of KEV-type vulnerabilities than the broader exploitation dataset used in previous research.

Our EPSS results show more variation than previous research, with efficiency of 4.9\% compared to the established 45.5\%. This variation highlights the importance of dataset characteristics and suggests that EPSS performance may be sensitive to the specific types of exploitation being measured. Most importantly, our study provides the first empirical evidence for the value of systematically integrating multiple vulnerability management approaches, showing that combined methods can achieve efficiency improvements while maintaining high coverage levels.

\subsection{Performance Stability and Practical Implementation}

The consistency of our main findings across different analytical perspectives provides evidence for the robustness of our approach. Our method maintains 85+ percent coverage while achieving substantial efficiency improvements, suggesting that performance is not dependent on specific data collection methodologies. While absolute efficiency values vary between analytical approaches, the relative performance rankings remain consistent, with our integrated approach outperforming individual methods while maintaining balanced efficiency-coverage trade-offs.

We conducted sensitivity analysis around our chosen EPSS threshold of 0.088 to validate the robustness of our parameter selection. Testing thresholds from 0.05 to 0.15 showed that efficiency decreases and coverage increases as the threshold is lowered, with the 0.088 threshold providing optimal balance based on the efficiency-coverage trade-off curve. Similarly, testing CVSS thresholds from 6.0 to 8.0 confirmed that 7.0 provides appropriate balance between workload reduction and risk acceptance.

\section{Discussion}
\label{sec:discussion}

This section interprets our experimental findings, discusses their implications for vulnerability management practice, addresses limitations of our approach, and identifies directions for future research. Our results provide strong evidence for the practical value of integrating multiple vulnerability management frameworks while revealing important areas for continued development.

\subsection{Key Findings and Implications}

Our central hypothesis that combining CVSS, EPSS, and KEV would achieve better efficiency while maintaining coverage receives strong empirical support. The substantial efficiency improvements over traditional CVSS-based approaches, coupled with maintenance of 85+ percent coverage levels, demonstrate that systematic integration of multiple data sources can overcome the limitations inherent in any single approach.

These results support the broader principle that cybersecurity decision-making benefits from multi-source intelligence integration rather than reliance on individual metrics. The complementary nature of historical exploitation evidence (KEV) and predictive modeling (EPSS) validates threat intelligence frameworks that emphasize diverse data source integration. For organizations currently struggling with CVSS-based vulnerability management, our approach provides immediate operational advantages. The approximately 95\% reduction in urgent prioritization workload from 16,182 to approximately 850 vulnerabilities while maintaining comprehensive coverage represents a transformative operational improvement.

KEV's exceptional performance on vendor report data (74.3\% efficiency, 86.7\% coverage) suggests strong alignment between CISA's threat intelligence sources and commercial security vendor reporting. This finding validates KEV as particularly valuable for organizations that rely heavily on vendor threat intelligence.

EPSS shows more consistent but generally lower efficiency across our analysis compared to previous research. This variation highlights the importance of understanding that machine learning-driven prediction models may perform differently across various organizational contexts and threat landscapes. Importantly, our integrated approach maintains stable performance characteristics, suggesting that the framework is robust to different types of exploitation evidence and organizational monitoring capabilities.

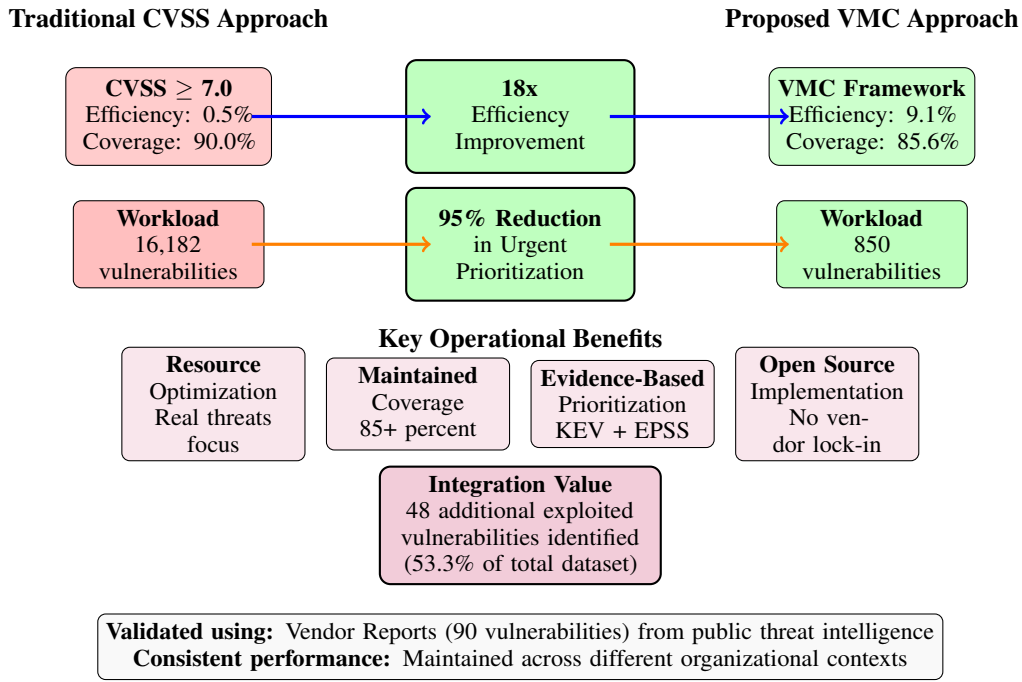
\begin{figure*}[t]
\centering
\begin{tikzpicture}[
    scale=0.85,
    every node/.style={font=\footnotesize},
    title/.style={font=\large\bfseries},
    subtitle/.style={font=\normalsize\bfseries},
    metric/.style={rectangle, draw, fill=blue!10, text width=2.5cm, text centered, minimum height=1.3cm, rounded corners=4pt},
    improvement/.style={rectangle, draw, fill=green!25, text width=2.8cm, text centered, minimum height=1.5cm, rounded corners=4pt, thick},
    workload/.style={rectangle, draw, fill=orange!15, text width=2.3cm, text centered, minimum height=1cm, rounded corners=3pt},
    benefit/.style={rectangle, draw, fill=purple!10, text width=2.2cm, text centered, minimum height=0.8cm, rounded corners=3pt}
]

\node[title] at (0, 9) {Vulnerability Management Chaining: Efficiency Improvements};

\node[subtitle] at (-5.5, 7.5) {Traditional CVSS Approach};
\node[metric, fill=red!20] at (-5.5, 6) {\textbf{CVSS $\geq$ 7.0}\\Efficiency: 0.5\%\\Coverage: 90.0\%};
\node[workload, fill=red!25] at (-5.5, 4) {\textbf{Workload}\\16,182\\vulnerabilities};

\node[subtitle] at (5.5, 7.5) {Proposed VMC Approach};
\node[metric, fill=green!20] at (5.5, 6) {\textbf{VMC Framework}\\Efficiency: 9.1\%\\Coverage: 85.6\%};
\node[workload, fill=green!25] at (5.5, 4) {\textbf{Workload}\\850\\vulnerabilities};

\node[improvement] at (0, 6) {\textbf{18x}\\Efficiency\\Improvement};
\node[improvement] at (0, 4) {\textbf{95\% Reduction}\\in Urgent\\Prioritization};

\draw[->, very thick, blue] (-4.2, 6) -- (-1.4, 6);
\draw[->, very thick, blue] (1.4, 6) -- (4.2, 6);
\draw[->, very thick, orange] (-4.2, 4) -- (-1.4, 4);
\draw[->, very thick, orange] (1.4, 4) -- (4.2, 4);

\node[subtitle] at (0, 2.5) {Key Operational Benefits};

\node[benefit] at (-4.8, 1.5) {\textbf{Resource}\\Optimization\\Real threats focus};
\node[benefit] at (-1.6, 1.5) {\textbf{Maintained}\\Coverage\\85+ percent};
\node[benefit] at (1.6, 1.5) {\textbf{Evidence-Based}\\Prioritization\\KEV + EPSS};
\node[benefit] at (4.8, 1.5) {\textbf{Open Source}\\Implementation\\No vendor lock-in};

\node[improvement, fill=purple!20, text width=3.5cm] at (0, -0.4) {\textbf{Integration Value}\\48 additional exploited\\vulnerabilities identified\\(53.3\% of total dataset)};

\node[draw, fill=gray!5, text width=11cm, text centered, minimum height=0.8cm, rounded corners=3pt] at (0, -2.3) {
    \textbf{Validated using:} Vendor Reports (90 vulnerabilities) from public threat intelligence\\
    \textbf{Consistent performance:} Maintained across different organizational contexts
};

\end{tikzpicture}
\caption{Operational efficiency improvements achieved by Vulnerability Management Chaining compared to traditional CVSS-based prioritization. The framework achieves 18x efficiency improvement while maintaining 85.6\% coverage, reducing urgent prioritization workload by 95\% (from 16,182 to ~850 vulnerabilities). Integration of KEV and EPSS identifies 48 additional exploited vulnerabilities (53.3\% of dataset) that neither method would capture individually.}
\label{fig:efficiency_improvements}
\end{figure*}

Figure~\ref{fig:efficiency_improvements} provides a comprehensive visualization of the operational improvements achieved by our Vulnerability Management Chaining framework. The dramatic efficiency gains and workload reductions demonstrated across multiple datasets highlight the transformative potential of systematic threat intelligence integration for organizational vulnerability management practices.

\subsection{Practical Implementation Guidance}

Organizations implementing our framework should consider several practical factors to maximize effectiveness. Initial deployment can begin immediately using existing tools and processes. Most vulnerability scanners already collect CVSS scores, requiring only the addition of KEV and EPSS data lookups. Simple scripts or spreadsheet formulas can implement the decision logic without requiring specialized security orchestration platforms.

Operational integration should align with existing patch management cycles. Critical Priority vulnerabilities (KEV-listed with CVSS $\geq$ 7.0) warrant emergency change procedures, while High Priority vulnerabilities can follow accelerated but planned maintenance windows. The Monitor category provides valuable input for threat hunting and detection engineering teams who can develop specific monitoring for exploited but lower-severity vulnerabilities.

Resource allocation becomes more strategic with clear prioritization tiers. Security teams can confidently defer approximately 95\% of vulnerabilities to standard patching cycles, focusing specialized resources on the ~5\% with evidence of exploitation threat. This dramatic workload reduction enables deeper analysis of truly critical vulnerabilities, including understanding attack patterns, validating compensating controls, and ensuring complete remediation.

\subsection{Evolution of Threat Intelligence Landscape}

Since our data collection period, the vulnerability management landscape has continued to evolve rapidly. EPSS models undergo regular retraining with improved feature engineering and expanded threat intelligence feeds. KEV catalog growth has accelerated, with CISA adding vulnerabilities more frequently as threat intelligence sharing improves. Commercial vendors have introduced new prioritization frameworks attempting to address limitations we identify.

These developments underscore both the timeliness of our research question and the industry-wide recognition that single-metric vulnerability management approaches are fundamentally insufficient for modern threat environments.

Rather than limiting the relevance of our findings, these advances validate our core premise: systematic integration of complementary intelligence sources provides superior vulnerability prioritization compared to any individual metric alone. Our framework establishes methodological principles that transcend specific tool implementations. The decision tree structure is designed to accommodate new intelligence sources, evolving prediction models, and adjusted threshold values without requiring fundamental architectural changes. This adaptability ensures that the integration methodology remains applicable as the threat intelligence ecosystem continues to evolve, providing organizations with a stable framework for incorporating emerging capabilities while maintaining operational consistency and proven effectiveness.

Several factors may limit the generalizability of our specific findings. Our exploitation dataset, while comprehensive, represents specific organizational contexts and time periods. Performance may vary in different threat environments or against different attack patterns. Vulnerability management performance may change over time as threat actor behaviors evolve, new exploitation techniques emerge, or defensive capabilities improve. Different organizations face different threat profiles based on their industry, size, geographic location, and security posture.

Our current approach performs static analysis of vulnerability characteristics without considering dynamic factors such as asset criticality, network exposure, or organizational-specific threat intelligence. The decision tree structure requires binary choices at each decision point, potentially losing nuanced information that could inform more sophisticated prioritization decisions. While our framework allows threshold adjustment, it provides limited mechanisms for incorporating organization-specific knowledge or contextual factors.

Several near-term improvements could enhance the framework's effectiveness. Machine learning approaches could automatically adjust EPSS and CVSS thresholds based on organizational feedback and historical performance data. Incorporating asset criticality and network exposure information could provide more sophisticated risk assessment while maintaining operational simplicity. Formal decision analysis techniques could replace simple threshold-based decisions with more nuanced multi-factor assessment while preserving automation potential.

Extended studies tracking framework performance over multiple years could identify temporal patterns and inform adaptive management strategies. Research examining how different types of organizations adapt and customize the framework could inform best practices and implementation guidance. Investigation of how emerging threat intelligence sources could be incorporated while maintaining the framework's simplicity and effectiveness would ensure continued relevance as the threat intelligence ecosystem evolves.

\section{Conclusion}
\label{sec:conclusion}

This research addresses a critical challenge in modern cybersecurity: the overwhelming volume of vulnerabilities that organizations must evaluate and prioritize for remediation. Through systematic integration of complementary vulnerability intelligence sources, we demonstrate that dramatic efficiency improvements are achievable while maintaining comprehensive security coverage.

\subsection{Summary of Contributions}

Our Vulnerability Management Chaining framework represents the first systematic methodology for integrating CVSS technical severity assessment, EPSS predictive modeling, and KEV confirmed exploitation evidence into a unified decision framework. This integration addresses fundamental limitations in each individual approach while preserving their unique strengths. Our research demonstrates that systematic integration of multiple data sources can overcome the fundamental limitations that plague individual vulnerability management approaches.

We developed the first systematic methodological framework for combining CVSS technical severity assessment, EPSS predictive threat intelligence, and KEV confirmed exploitation evidence, establishing foundational principles for multi-source vulnerability intelligence integration that transcend specific tool implementations. Using 28,377 vulnerabilities and exploitation evidence from multiple sources, we demonstrated significant efficiency improvements over traditional CVSS-based prioritization while maintaining 85+ percent coverage of actually exploited vulnerabilities.

Our analysis reveals that KEV and EPSS identify complementary sets of exploited vulnerabilities, with their combination capturing 48 additional vulnerabilities that neither method would identify individually. By relying exclusively on freely available data sources, our framework enables broad adoption without requiring expensive commercial threat intelligence subscriptions.

Organizations adopting our framework can reduce their urgent vulnerability remediation workload by approximately 95\% while capturing 85+ percent of vulnerabilities that attackers actually exploit. This substantial workload reduction enables security teams to allocate resources more effectively and focus on truly critical threats. Unlike complex alternatives requiring specialized expertise, our decision tree framework can be implemented using standard vulnerability management tools and provides immediate operational benefits for organizations regardless of their size or security maturity level.

\subsection{Future Directions and Final Remarks}

As the cybersecurity landscape continues to evolve, our methodological framework provides a foundation for incorporating emerging threat intelligence sources and enhanced prediction models. The demonstrated effectiveness of systematic multi-source integration suggests that future developments should focus on adaptive frameworks that can seamlessly incorporate new intelligence feeds while maintaining operational simplicity and proven effectiveness.

The exponential growth in vulnerability disclosures makes efficient vulnerability management increasingly critical for organizational security. Our Vulnerability Management Chaining framework provides a practical solution that leverages existing open source data to achieve dramatic efficiency improvements while maintaining comprehensive security coverage. More broadly, our research demonstrates that systematic integration of multiple intelligence sources can overcome the limitations inherent in individual approaches, providing a methodology applicable beyond vulnerability management to other cybersecurity domains.

We encourage security practitioners to implement and adapt our framework to their specific organizational contexts, and we invite researchers to build upon our methodology to develop even more effective approaches to this critical challenge. Only through continued innovation in vulnerability management can defenders hope to keep pace with the ever-expanding threat landscape.


\bibliographystyle{IEEEtran}
\bibliography{references}  






\end{document}